\documentclass[reprint,aps,prl,superscriptaddress,showpacs]{revtex4-1}
\usepackage{graphicx}
\usepackage{dcolumn}
\usepackage{bm}
\usepackage{color}
\usepackage{soul}
\usepackage{amsmath}
\usepackage[english]{babel}

\definecolor{AFB}{rgb}{0.8, 0.3, 0.2}

\begin{document}

\title{Statistical properties of nonlinear stage of modulation instability in fiber optics}

\author{Adrien Kraych}

\affiliation{Laboratoire de Physique des Lasers, Atomes et Molecules,  UMR-CNRS 8523,  Universit\'e de Lille, France}
\affiliation{ Centre d'Etudes et de Recherches Lasers et Applications (CERLA), 59655 Villeneuve d'Ascq, France}

\author{Dmitry Agafontsev}

\affiliation{ P.P. Shirshov Institute of Oceanology, 36 Nakhimovsky prosp., Moscow 117218, Russia}

\affiliation{Novosibirsk State University, Novosibirsk, 630090, Russia}

\author{St\'ephane Randoux}
\affiliation{Laboratoire de Physique des Lasers, Atomes et Molecules,  UMR-CNRS 8523,  Universit\'e de Lille, France}
\affiliation{ Centre d'Etudes et de Recherches Lasers et Applications (CERLA), 59655 Villeneuve d'Ascq, France}

\author{Pierre Suret} 
\affiliation{Laboratoire de Physique des Lasers, Atomes et Molecules,  UMR-CNRS 8523,  Universit\'e de Lille, France}
\affiliation{ Centre d'Etudes et de Recherches Lasers et Applications (CERLA), 59655 Villeneuve d'Ascq, France}

\email[Corresponding author : ]{Pierre.Suret@univ-lille.fr}

\begin{abstract}
  We present an optical fiber experiment in which we examine the space-time evolution of a modulationally unstable  plane wave initially perturbed by a small noise. Using a recirculating fiber loop as experimental platform, we report the single-shot observation of the noise-driven development of breather structures from the early stage to the long-term evolution of modulation instability.
Performing single-point statistical analysis of optical power recorded in the experiments, we observe decaying oscillations of the second-order moment together with the exponential distribution  in the long term evolution, as predicted in [D.\,S. Agafontsev and V.\,E. Zakharov, Nonlinearity {\bf 28}, 2791 (2015)]. Finally, we demonstrate experimentally and numerically that the autocorrelation of the optical power $g^{(2)}(\tau)$ exhibits some unique oscillatory features typifying the nonlinear stage of the noise-driven modulation instability and of integrable turbulence.
\end{abstract}

\maketitle

 {\it Modulational Instability} (MI), also known as {\it Benjamin-Feir Instability}, leads to the spontaneous breakup of plane waves into train of pulses. MI has been extensively studied in the last five decades, see e.g.~\cite{Benjamin:67,Zakharov:MI:09}, and is often seen as the exponential amplification of a weak modulation of a monochromatic carrier wave. This fundamental phenomenon  arises in physical systems such as deep water waves~\cite{Osborne},  nonlinear optical waves~\cite{Agrawal}, Bose-Einstein condensates and matter waves~\cite{Strecker:02}, which are described at leading order by the focusing one-dimensional nonlinear Schr\"odinger equation (1DNLSE)~\cite{Zakharov:MI:09}. 

The 1DNLSE is integrable and its exact solutions can be found by using the so-called ``inverse scattering transform'' (IST)~\cite{Shabat:72,Ablowitz:96}. For the MI driven by a sinusoidal perturbation of a constant background, the long-term spatio-temporal dynamics is described by {\it breather} solutions of the focusing 1DNLSE, such as the Akhmediev Breather (AB)~\cite{Akhmediev:85,Akhmediev:86,Akhmediev:09, Chabchoub:11,Kibler:12,Kibler:15, Pierangeli:18, Mussot:18}.  However, natural perturbations are generally not sinusoidal modulations but rather represent a random process~\cite{Toenger:15,Dudley:14,Akhmediev:09b,Agafontsev:15,Solli:12,Randoux:IST:16,Narhi:16}.

Despite the numerous studies devoted to the subject of MI in optical fibers~\cite{Agrawal,Tai:86}, it is only recently that single-shot observations of the noise driven MI have  been reported. Shot-to-shot fluctuations of the optical spectra have been measured in experiments with light pulses in~\cite{Solli:12}, while breather structures spontaneously emerging from noise superimposed to a continuous wave (CW) field have been reported in \cite{Narhi:16}.  However, in these pioneering studies the spatio-temporal dynamics and the evolution of statistical properties of the light field along the fiber could not be recorded, since the observation  was made only at the output of the fiber.

From the theoretical point of view, the question of noise-driven MI enters whithin the fundamental framework of {\it integrable turbulence}, which was first introduced by V.\,E. Zakharov in~\cite{Zakharov:09} and is now a
subject of extensive theoretical~\cite{Zakharov:09,Zakharov:13,Agafontsev:14,Agafontsev:15,Agafontsev:16,SotoCrespo:16,Akhmediev:IntegrableTurbulence:16,Onorato:16} and experimental~\cite{Walczak:15,Suret:16,Randoux:16,Randoux:17,Koussaifi:18,Tikan:18} research. In the long term evolution, the integrable turbulence is characterized by a steady state where statistical properties become invariant. Surprisingly, for initial conditions made of a plane wave with small additional noise, numerical simulations of the focusing 1DNLSE show that the long-term single-point probability density function (PDF) of the field $\psi$ is Gaussian while the PDF of the  power $P=|\psi|^2$ is exponential~\cite{Agafontsev:15,Akhmediev:IntegrableTurbulence:16,SotoCrespo:16}.  Moreover, the transient regime is characterized by the decaying oscillatory evolution of the statistical moments~\cite{Agafontsev:15}.  To the best of our knowledge, up to now, this remarkable transient regime has not been observed in experiments.

In this Letter, we investigate the evolution of a modulationally unstable plane wave   by using a recirculating fiber loop \cite{Lacourt:02, Kraych:19}. Our setup enables the single-shot recording both in space and time of the nonlinear dynamics associated with the noise-driven MI. The spatio-temporal dynamics observed in the experiments is very similar to the one previousy predicted by numerical simulations, see e.g.~\cite{Toenger:15, Dudley:14}.  The experimental data analysis also reveal the decaying oscillations of the statistical moments predicted in~\cite{Agafontsev:15}. 
Moreover, we demonstrate numerically and experimentally that the autocorrelation of optical power $g^{(2)}(\tau)$ exhibits a clear oscillatory signature that allows to differentiate the wavefield in the nonlinear stage of the MI from random Gaussian fields made up of linear superposition of waves.

\begin{figure}[t]
\includegraphics[width=8cm]{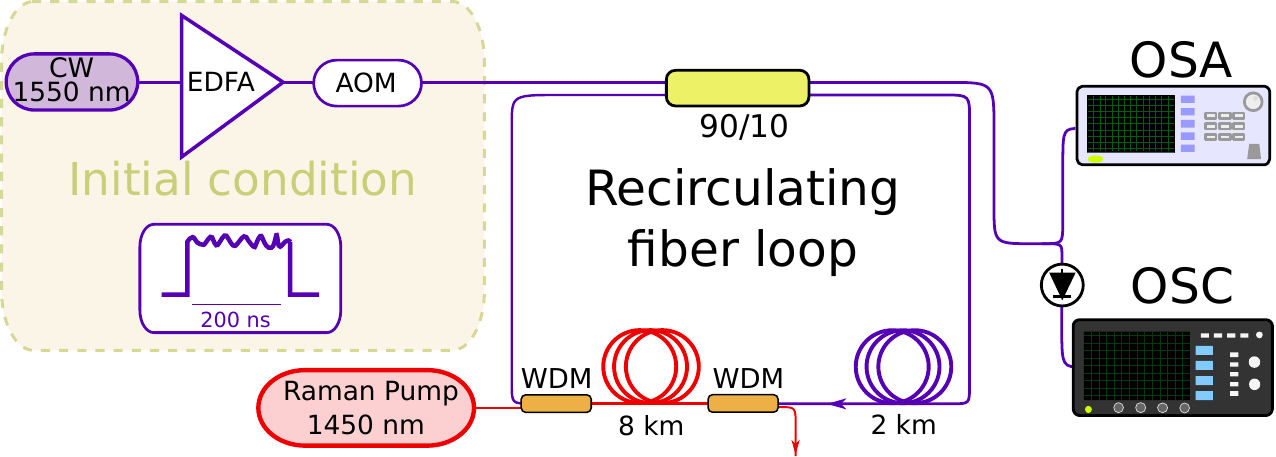}

\caption{{\bf Schematic experimental setup.} 
 The initial field is emitted by a single frequency laser source (CW) amplified
by an EDFA. An acousto-optical modulator (AOM) allows to inject
a 200-ns pulse inside the 10-km long recirculating fiber loop.
The losses of the cavity are partly compensated by  Raman
amplification in a 8-km long section of the loop. The signal
is observed at each round trip by using ultrafast photodiode, oscilloscope (OSC) and an optical spectrum analyser (OSA)
}
\label{fig:1}
\end{figure}

Our experimental setup is very similar to the one used in~\cite{Kraych:19} and is schematically shown in Fig.~\ref{fig:1}. It consists of a recirculating fiber loop, {\it i.e.} a  ring cavity made up of $10$~km single mode fiber (SMF) closed on itself by a $90/10$ fiber coupler (FC). A square-shape light pulse having a temporal width around $200$~ns is first prepared by using a single frequency laser operating at $1550$~nm, an Erbium-doped fiber laser (EDFA) and an acousto-optical modulator (AOM) (see~\cite{Kraych:19} and Supplementary Material for details). Secondly, this long square-pulse is launched into the FC which is arranged in such a way that $90\%$ of the power is recirculated inside the fiber loop.

The square-shape pulse plays the role of the initial condensate (plane wave) perturbed by the noise naturally added by the spontaneous emission of the EDFA. 
Pulses are injected inside the loop with a period of $10$~ms, which is much larger than the cavity round-trip time of $49$~$\mu$s. 
The linear losses of the SMF loop are partially compensated by Raman amplification in a $8$~km long section of the loop. Following the method used in~\cite{Mussot:18, Kraych:19}, the Raman gain is provided by a Raman pump operating at $1450$~nm and launched in a counter-propagating direction. 
Our setup enables propagation of the signal over an effective propagation length $1/\alpha_{\rm eff}\simeq 600$~km, where $\alpha_{\rm eff}\simeq 1.7\times 10^{-3}$~km$^{-1}$ is the effective losses coefficient measured from the decay rate of the signal inside the ring cavity (see Supplementary Material). 
Note that the standard linear losses coefficient of the SMF (without compensating amplification) is $46\times 10^{-3}$~km$^{-1}$ (equivalently $0.2$~dB/km).

Our experiment is designed to propagate light fields  having low optical power  (tens of milliwatts)  over long distances, enabling the development of the nonlinear stage of MI with time scales of tens of picoseconds. The signal is thus simply recorded by using fast photodiode connected to a fast oscilloscope. The global detection bandwidth evaluated by using a first-order low-pass filter model is $28$~GHz, whereas the theoretical frequency of the maximum gain of the MI is $\nu_{M} \simeq 12$~GHz (see Supplementary Material). Note that we have checked that stimulated Brillouin scattering does not affect our observations.

\begin{figure*}[t]
\includegraphics[width=0.9\linewidth]{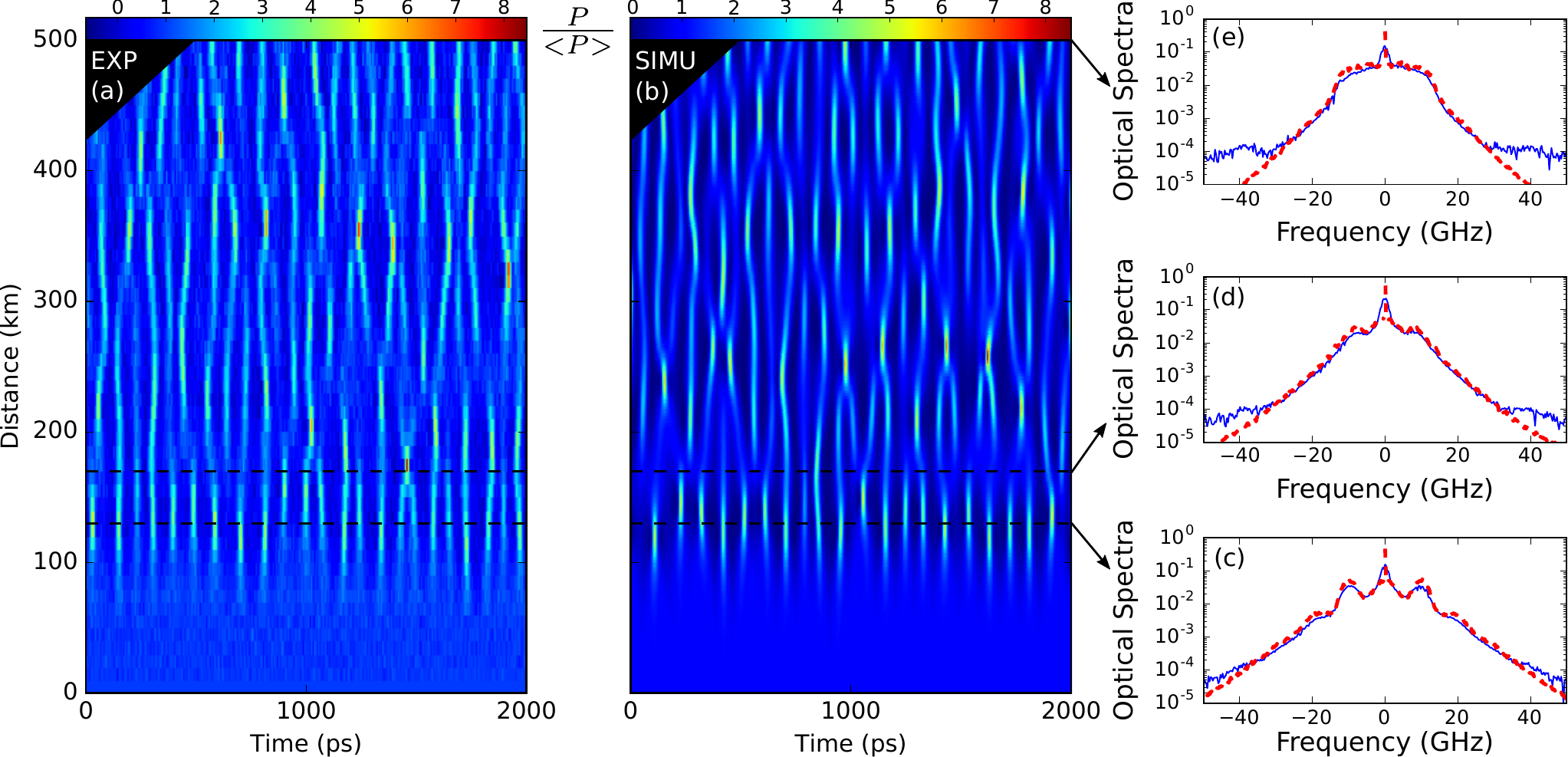}

\caption{{\bf Typical space-time evolution of the spontaneous modulation
    instability} {\bf (a)} Experiment {\bf (b)} Numerical simulation
  of 1DNLSE (\ref{eq:NLSE}) {\bf (c-e) Optical spectra}  recorded in experiments (blue lines) and computed from numerical  simulations (dashed red lines)   for different propagation distances  $z$: (c)  $z=130$ km, (d)  $z=170$ km,  (e)  $z=500$ km. The parameters used in simulations of Eq. (\ref{eq:NLSE}) are $\beta_2=-22$ ps$^2$/km, $\gamma=1.3$ W$^{-1}$km$^{-1}$, mean optical power of the initial field $P_0=43$mW and  $\alpha_{\rm eff} \simeq 1.7 \times 10^{-3}$ km$^{-1}$.}
\label{fig:2}
\end{figure*}

Fig.~\ref{fig:2}(a) displays the typical space-time evolution of the optical power recorded in single-shot in our experiments. In the figure, a $2$~ns-wide portion of the $200$~ns circulating pulse has been selected to show the space-time dynamics over a total propagation distance of $500$~km (see Supplementary Material for details). In Fig. 2(a), one can easily recognize a typical pattern of the noise-driven MI reported in various numerical studies, see e.g.~\cite{Dudley:14,Toenger:15}. 

Experiments presented in this Letter are well described by the numerical simulation of the 1DNLSE with an additional term describing  the effective power losses~\cite{Kraych:19}, 
\begin{equation}\label{eq:NLSE}
i\frac{\partial \psi}{\partial z}=\frac{\beta_2}{2}\frac{\partial^2
  \psi}{\partial t^2}-\gamma|\psi|^2\psi -i \frac{\alpha_{\rm eff}}{2}
\psi.
\end{equation}
Here $\psi(z,t)$ represents the complex envelope of the electric field that slowly varies in space $z$ and time $t$. 
At $1550$~nm, the group velocity dispersion coefficient of the SMF is $\beta_2=-22$~ps$^2$/km and the Kerr coefficient is $\gamma=1.3$~W$^{-1}$km$^{-1}$. The optical power of the initial signal measured in experiments is $P_{0} \approx 48$~mW and the largest propagation distance  $z_{max}=500$~km  corresponds to an effective  propagation length  $L_{\rm eff}/L_{\rm nl}\sim 21$, where $L_{\rm  eff}=[1-\exp(-\alpha_{\rm eff} z_{max})]/\alpha_{\rm eff}$ and $L_{\rm  nl}=1/(\gamma P_0)$  is the nonlinear length scale, see e.g.~\cite{Agrawal}. Note that in all simulations the power is slightly adjusted to $P_0=43$ mW (see Supplementary).

Fig.~\ref{fig:2}(b) represents the typical spatio-temporal dynamics obtained in numerical simulations of Eq.~(\ref{eq:NLSE}) and shows a remarkable qualitative agreement with the typical dynamics recorded in the experiment [Fig.~\ref{fig:2}(a)]. In particular, in the early nonlinear stage of the noise-driven MI, one can notice the emergence of the well-known quasi-periodic structures  resembling the Akhmediev breathers~\cite{Toenger:15,Akhmediev:86}. Figs.~\ref{fig:2}(c-e) demonstrate quantitative agreement between the numerical and the experimental optical spectra. Let us finally emphasize that the observed dynamics is only slightly perturbed by the weak  losses experienced by the light wave over the propagation distance of $500$ km (see the Supplementary Material).

We now examine the evolution of the statistical properties of the waves during the propagation. Note that, as we consider  statistical processes independent of time $t$, statistical distributions and moments  are computed by averaging both over time $t$ and over ensemble of realizations. In the following experimental data analysis, statistical variables are computed  by averaging the results over $\sim 15000$ points extracted from the central part of $19$ different $200$~ns-wide pulses recorded  at a sampling rate of $160$~Gs/s.

\begin{figure*}[t]
\includegraphics[width=0.9\linewidth]{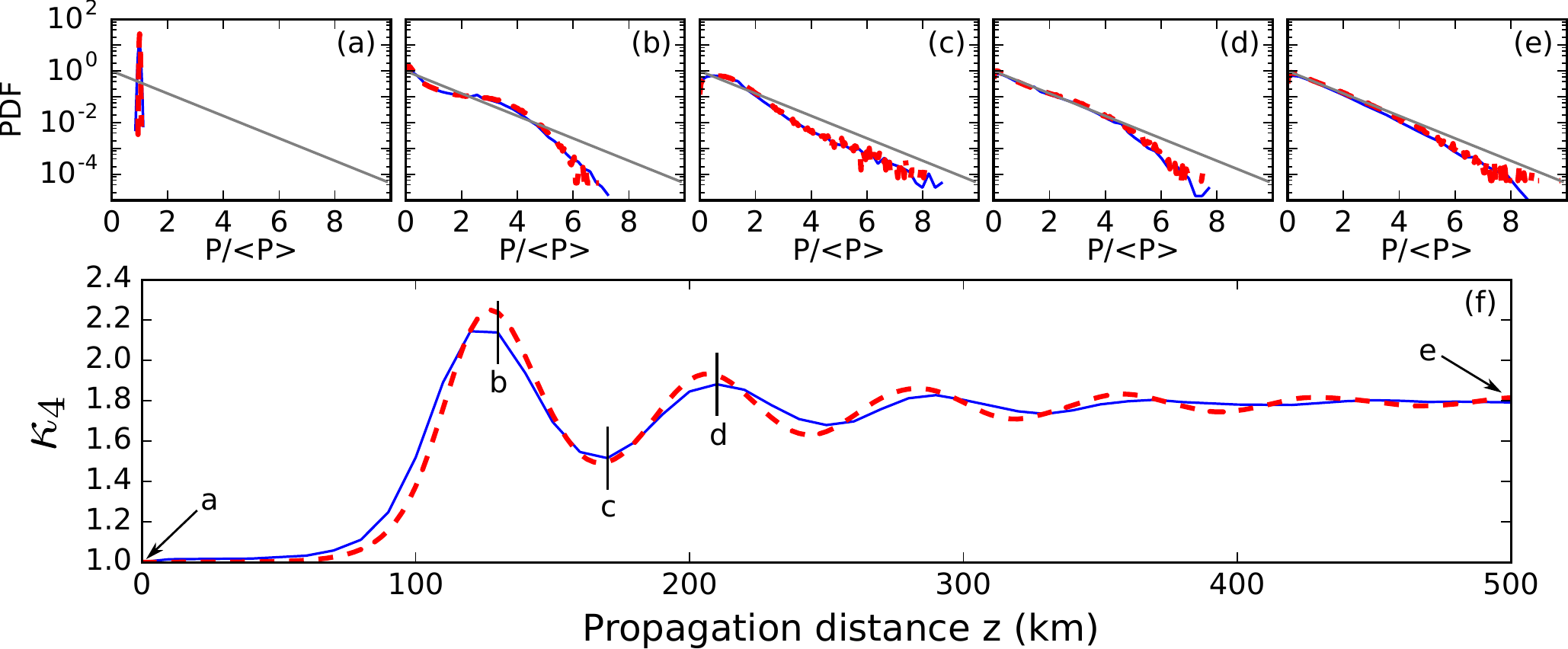}

\caption{{\bf Single point statistics of optical power} {(a-e)
    Probability  density functions (PDF) of the normalized optical
    power} $P/\langle P \rangle$ for different propagation distances  (a) $z=0$~km,   (b) $z=130$~km, (c)  $z=170$~km, (d)  $z=210$~km and (e)   $z=500$~km.  The exponential distribution is plotted with thin grey lines.  (f)~Evolution of the second order moment  $\kappa_4(z)=   \langle  P(z,t)^2  \rangle / \langle P(z,t)  \rangle^2$ of the  optical power with propagation distance.  Experimental results are plotted in blue lines  and numerical simulations of  1DNLSE  (Eq. (\ref{eq:NLSE}), with $\beta_2=-22$ ps$^2$/km,  $\gamma=1.3$W$^{-1}$km$^{-1}$, mean initial power $P_0=43$~mW and  $\alpha_{\rm eff} \simeq 1.7 \times 10^{-3}$ km$^{-1}$) are plotted in dashed red lines. Finite detection bandwidth has been  included in the simulation.}
\label{fig:3}
\end{figure*}

First,  we compute the PDF of the optical power of the signal recorded at each round trip in the fiber loop. The PDF measured at  different  propagation distances $z$ is displayed with blue lines in Figs.~\ref{fig:3}(a-e). As expected, for the initial slightly perturbed plane wave, the distribution is a narrow peak centered around the average power  [see Fig.~\ref{fig:3}(a)].  For larger $z$, the  emergence of breather-like structures observed in the space-time evolution (Fig.~\ref{fig:2}(a))  yields     significant  changes of the PDF, (see  Figs.~\ref{fig:3} (b-e)). In particular, one sees the emergence of oscillatory tails [see Figs.~\ref{fig:3}(b-e)], qualitatively very similar to those computed numerically for the integrable 1DNLSE ($\alpha_{\rm eff}=0$ in Eq.~(\ref{eq:NLSE})) in~\cite{Agafontsev:15}. At long evolution distance, the experimental PDF is close to the exponential distribution [thin grey lines in Figs.~\ref{fig:3}(a-e)], which is also predicted for the integrable case~\cite{Agafontsev:15}.

The small difference between the exponential distribution and the measured PDF in Fig.~\ref{fig:3}(e) arise from the limited detection bandwidth $\sim 30$GHz of the photodiode. By taking into account this effect in  simulations, the numerical PDFs [dashed red lines in Fig.~\ref{fig:3}(a-e)] are in quantitative agreement with the experimental PDFs (see Supplementary Material).

Figures~\ref{fig:3}(a-e) reveal a nontrivial evolution of the PDF along the optical fiber. One can describe this evolution with a single statistical indicator by using the normalized fourth-order moment of the amplitude $|\psi|$, which is also the second-order moment of the optical power $P=|\psi|^2$,
\begin{equation}
\kappa_4(z)=\frac{\langle P(z,t)^2\rangle}{\langle P(z,t)\rangle^2}.
\label{eq:kappa4}
\end{equation}
Note that  $\kappa_4$ is proportional to the nonlinear part of the Hamiltonian of the 1DNLSE, see e.g.~\cite{Agafontsev:15,Walczak:15,Onorato:16}.   The
blue line in Fig.~\ref{fig:3}(f) represents  the evolution of  $\kappa_4(z)$
computed from  experimental data recorded every $10$ km in the fiber loop. The experimental second order moment exhibits a remarkable transient regime in the form of decaying
oscillations, qualitatively very similar to the one predicted
in~\cite{Agafontsev:15} from  numerical simulations of the integrable 1DNLSE [$\alpha_{\rm eff}=0$ in Eq.~(\ref{eq:NLSE})]. These oscillations of $\kappa_4(z)$ might be interpreted as a statistical signature of Fermi-Pasta-Ulam reccurences recently in optical experiments considering the evolution of a plane wave perturbed by a sine (deterministic) modulation~\cite{Mussot:18, Pierangeli:18}. Here, despite the randomness of the initial perturbation, we observe some form of (statistical) reccurences~\cite{WabnitzMI:14}.

The exponential PDF corresponds to the value of the second order moment $\kappa_4 = 2$, whereas at long propagation distance it only reaches $\kappa_4\simeq 1.8$ in our experiments. The difference  between the two values arises from the finite bandwidth of detection, as explained above.  By taking into account this effect in simulations, the numerically computed moment $\kappa_4(z)$ [dashed red line in Fig.~\ref{fig:3}(f)] is in quantitative  with the experimental one.  Note that the effective linear losses mainly influence  the spatial period of the oscillations of $\kappa_4$, but almost do not affect the stationary value (see Supplementary material).

\begin{figure*}[t]
\includegraphics[width=0.9\linewidth]{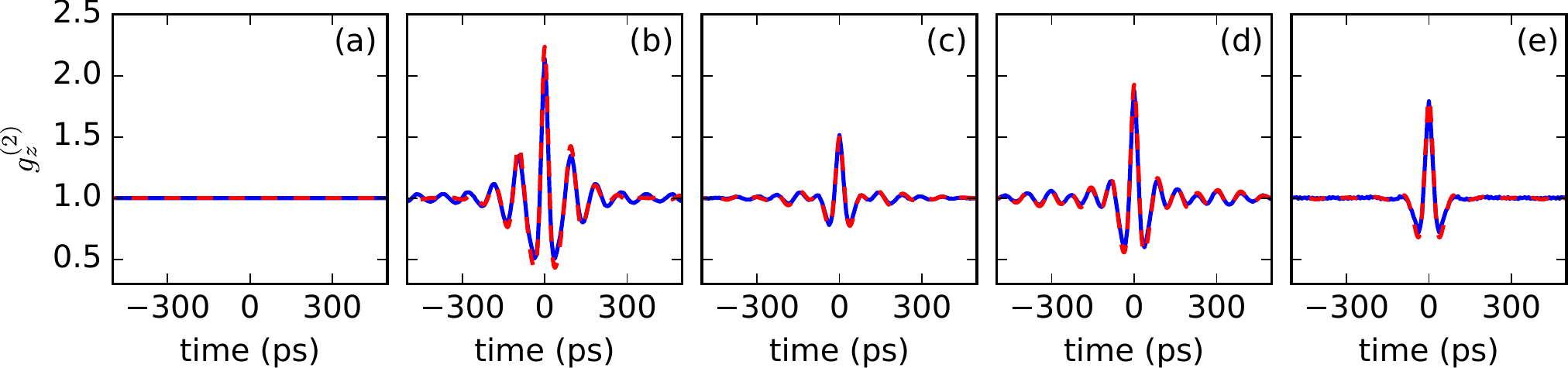}

\caption{ {\bf Double-points statistics of optical power.}  Autocorrelation of the power $g_z^{(2)}(\tau)$ [see Eq. (\ref{eq:g2})]  at propagation distances $z$ corresponding to those shown in Figs.~\ref{fig:3}(a-e): (a) $z=0$~km, (b) $z=130$~km, (c)  $z=170$~km, (d) $z=210$~km and (e) $z=500$~km.  Experimental results are plotted in blue lines, while numerical results are plotted in dashed red lines. Parameters of numerical simulations (including finite bandwidth of detection) are identical to those plotted in Fig. \ref{fig:3}.}
\label{fig:4}
\end{figure*}

The exponential distribution of the optical power $P=|\psi|^2$ characterizing the stationary state of  integrable turbulence~\cite{Agafontsev:15} is identical to the PDF characterizing a field $\psi$ made up from a linear superposition of large amount of waves with random phases~\cite{Walczak:15,Randoux:16}. 
In order to distinguish between these two processes  of profoundly different natures, we now  introduce the second-order degree of coherence of the field, i.e., the optical power autocorrelation  function defined for stationary processes (independent of time $t$)  as :
\begin{equation}
g_z^{(2)}(\tau)= \frac{\langle P(z,t)P(z,t+\tau)\rangle}{\langle P(z,t)\rangle^{2}}.
\label{eq:g2}
\end{equation}
The autocorrelation is an even function  and is equal to the second order moment  at $\tau=0$ [$g_z^{(2)}(0)=\kappa_4(z)$].  In the case of a pure Gaussian process (e.g. thermal light), it has the maximal value of $g_z^{(2)}(0)=2$ and is bounded from below, $g_z^{(2)}(\tau)\ge 1$, see e.g.~\cite{Goodman:statistics:15}. The blue lines in Figs.~\ref{fig:4}(a-e) represent  the function $g_z^{(2)}(\tau)$ computed from the experimental data recorded at different  propagation distances $z$.   As  expected for the initial slightly perturbed condensate at $z=0$~km, the autocorrelation function is almost unity, $g_z^{(2)}(\tau)\simeq 1$, see Fig.~\ref{fig:4}(a).  In the nonlinear stage of the MI, $g_z^{(2)}(\tau)$ is characterized by an oscillatory shape  having a period $\approx 95$ps approximately equal to the inverse of the maximum gain MI frequency, $1/\nu_M=84$~ps. 
Remarkably, the $g_z^{(2)}(\tau)$ function reveals that the quasi-periodicity of the MI process is preserved at long propagation distance,  even when the initial perturbation of the condensate is a purely random process.

Taking into account the effective losses and the finite detection bandwidth, the numerical simulations (dashed red lines in Fig.~\ref{fig:4}) are in quantitative agreement with the experiments. 
Our study reveals that the effective losses play a significant role on the visibility of the oscillatory
structure of $g_z^{(2)}(\tau)$.  For sufficiently large
distances of propagation $z\gtrsim 200$~km, this structure gradually disappears
both  in experiments and numerical simulations.  At $z=500$~km,
only one oscillation of $g_z^{(2)}(\tau)$ is visible, thus revealing a
decrease of the degree of coherence.  On the contrary, numerical
simulations performed with zero effective losses $\alpha_{\rm eff}=0$
demonstrate the existence of the distinctive  oscillatory shape for any large propagation distance (see Supplementary Material).  \\

In conclusion, we have investigated experimentally and numerically the statistical properties of the spontaneous noise-driven MI, both in the long-term evolution and in the transient regime. 
First, we have reported the single-shot measurement of the typical spatio-temporal dynamics of the MI in  optical fiber. 
Second, for the first time in a real physical system, we have observed the transient oscillations of the fourth-order moment predicted in~\cite{Agafontsev:15} while the power exhibits an exponential distribution at long propagation distance. Finally, we have shown that in the nonlinear stage of the MI, the autocorrelation $g_{z}^{(2)}(\tau)$ of the optical power is characterized by a remarkable oscillatory shape.   Even though our experiments demonstrate these basic features of the integrable turbulence qualitatively very well, we have also shown that weak linear damping induce a long-term loss of coherence on the nonlinear random field.

\section*{Supplementary Material : Statistical properties of nonlinear stage of modulation instability in fiber optics}

\section{Section I: Optical fiber experiments}

\subsection{Experimental Setup}

We provide here some details of the experimental setup (Fig. \ref{fig:1}). The light source used for generation of the plane wave is a single-frequency continuous-wave (CW) laser diode (APEX-AP3350A) centered at 1550 nm which delivers an optical power of a few mW. The power of the generated plane wave is amplified to the Watt-level by using an Erbium-doped fiber amplifier (EDFA). The spontaneous emission of the EDFA acts as some noise that is added to the monochromatic plane wave.

\begin{figure}[h]
\includegraphics[width=8cm]{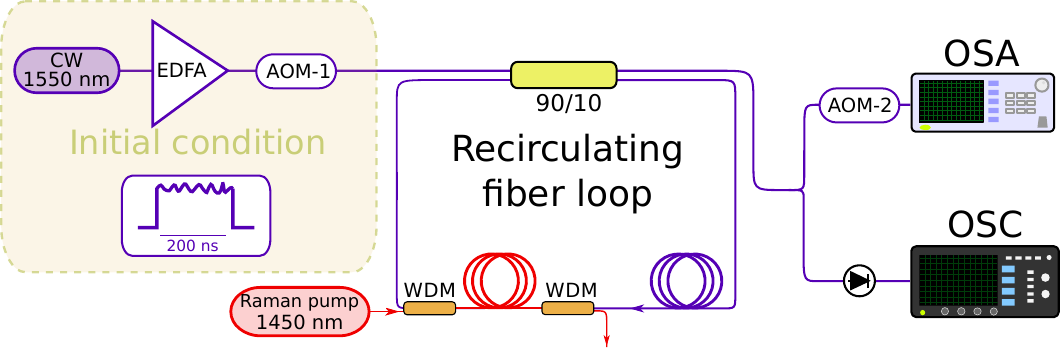}
\caption{{\bf Experimental setup} The initial field is emitted by a single frequency CW source amplified
by and Erbium doped fiber amplifier. An acousto-Optic modulator (AOM-1) allows to design 200-ns pulses which are launched inside the 10-km long recirculating fiber loop. The losses of the cavity are partly compensated by a Raman fiber amplifier (corresponding to 8 km of the fiber loop). The signal is observed at each round trip by using ultrafast photodiodes. The spectrum is observed by using optical spectrum analyzer (OSA): a second acousto-optical modulator (AOM-2) allows to select one specific pulse associated with a given propagation distance.}
\label{figSup:1}
\end{figure}

A periodic train of square pulses having a width of $ \approx$ 200 ns and a period of 10 ms is prepared by using an acousto-optic modulator (AOM-1). The duration of the  200~ns pulses  is much greater than the typical duration (30 - 50 ps) of the nonlinear structures that emerge from nonlinear propagation.

The recirculating fiber loop is made up of a ∼ 10 km of single-mode fiber (SMF) closed on itself by a 90/10 fiber coupler (FC). The FC is arranged in such a way that 90 $\%$ of the intracavity power is recirculated. The SMF has been manufactured by Draka-Prysmian. It has a measured second-order dispersion coefficient of $-22$ ps$^2$ km$^{-1}$ and an estimated Kerr coefficient of $1.3$ km$^{-1}$ W$^{-1}$ at the working wavelength of 1550 nm.

In all the experiments presented in this Letter, the measured optical power of the initial signal is $P_{0} = 48$~mW and the theoretical frequency associated with the maximum gain of the MI is $\nu_{M} = (1/2\pi)\sqrt{2 |\gamma P_0/\beta_2|}\simeq 12$~GHz. The accuracy of measurement of the power inside the cavity is around 10$\%$. In numerical simulations, we adjust slightly the power to $P_0=43$mW in order to match the period of the transient oscillations of $\kappa_4$ to the period measured in experiments. We believe that  this small difference betweeen the two values of $P_0$ arise from the use of a mean-field model where effective losses are introduced in a phenomenological way. Indeed, losses of the FC are localized at a discrete position inside the cavity and rigorously speaking this cannot be fully described by a simple model with effective losses, see Eq. (\ref{eq:NLSE}). It is remarkable that the effective model quantitatively reproduce experiments by a slight adjustment of the power. 

The optical signal at the output of the 90/10 FC is launched into a 50/50 FC, which allows to monitor simultaneously the dynamics and the spectrum of the nonlinear optical waves. First, the signal is recorded by using a fast photodiode (Picometrix D-8IR) connected to a fast oscilloscope (LeCroy LabMaster 10-65Zi). Secondly, an acousto-optic modulator (AOM-2) allows to select a specific round trip by adjusting the delay between an electrical gate that trigs AOM-2 after that a square pulse has been generated by AOM-1. The output light from AOM-2 is launched into an optical spectrum analyser (OSA,  YOKOGAWA - AQ6370)) allowing to monitor the optical power spectrum at a given propagation distance.

\subsection{Single-shot recording}

The Probability Density Function (PDF)  and the second order moment of the optical power are shown in Fig. 3 of the Letter. We used the fast oscilloscope in a sequence mode.

 As schematically shown in Fig. 2(a), the oscilloscope performs the
  periodic acquisition of the signal at a period of $10$ ms at which the $200$ ns
  square light pulses are also launched inside the recirculating fiber loop. As shown
  in Fig. 2(b), the signal is not recorded over the duration of $49$ $\mu$s
  that corresponds to the round trip time of light inside the fiber loop. It is recorded
  synchronously to the round trip time in a sequence mode where information is
  sampled over time windows that last $500$ ns. The sampling rate is equal
  to 160 Gs/s and each pulses of 200 ns is thus discretized by using $\simeq$ 32000 points.

\begin{figure}[h]
\includegraphics[width=9cm]{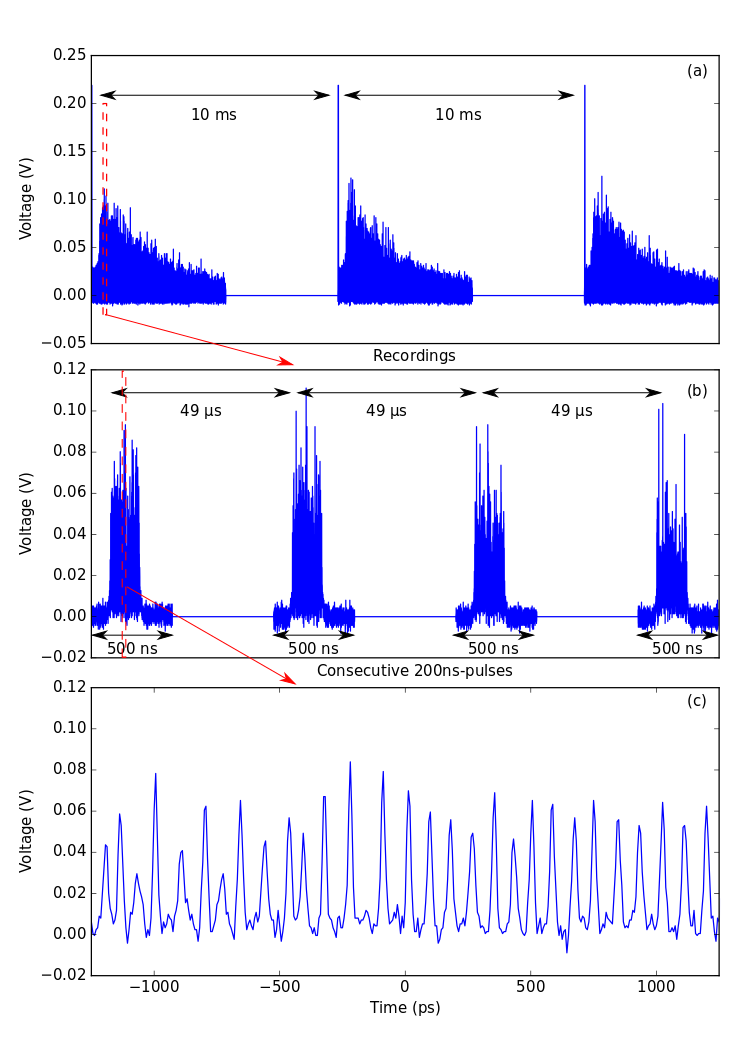}
\caption{ (a) Signal recorded by the fast photodiode at a period of $10$ ms
  fixed by the period at which light is launched inside the recirculating fiber loop.
  (b) Sequence of signals recorded by the oscilloscope evry $49$ $\mu$s. (c) Zoomed
  view of coherent structures recorded by the photodiode over$\simeq$ 2 ns (zoom)}
\label{figSup:2}
\end{figure}

  For the statistical analysis, we use only $15 000$ points that are extracted from
  the central part of the signal. This permits to clear possible measurement
  of any unwanted effect arising from the finite size of the square pulses.
  Statistical averages are finally computed from an ensemble $19$ pulses. Considering the time scale of structures ($\sim 100$ps), the statistical averaging is made typically over 20, 000 structures. Note that the emergence of extreme events induced by the modulation instability can be seend in Fig. \ref{figSup:2} (the maxima of instantaneous power roughly correspond to the first maximum of the second order moment studied in the Letter). On the contrary, the averaged energy at each round trip is slowly decaying (see above and Fig. \ref{figSup:2}).

\subsection{Raman amplification and effective losses}

Raman amplification is achieved by injecting a pump wave at a wavelength of 1450 nm inside the fiber loop. The pump wave is coupled in and out the recirculating fiber loop by using two commercial wavelength division multiplexers (WDM) splitting the1450nm and 1550 nm wavelengths. The length of the amplifying section is of 8 km. The pump laser at 1450 nm is a commercial Raman fiber laser delivering an output beam having a power of several Watt at 1450nm. In our experiments, this optical power is attenuated to typically $\approx$ 230 mW by using a 90/10 fiber coupler (not shown in Fig. \ref{figSup:1}).

\begin{figure}[h]
\includegraphics[width=9cm]{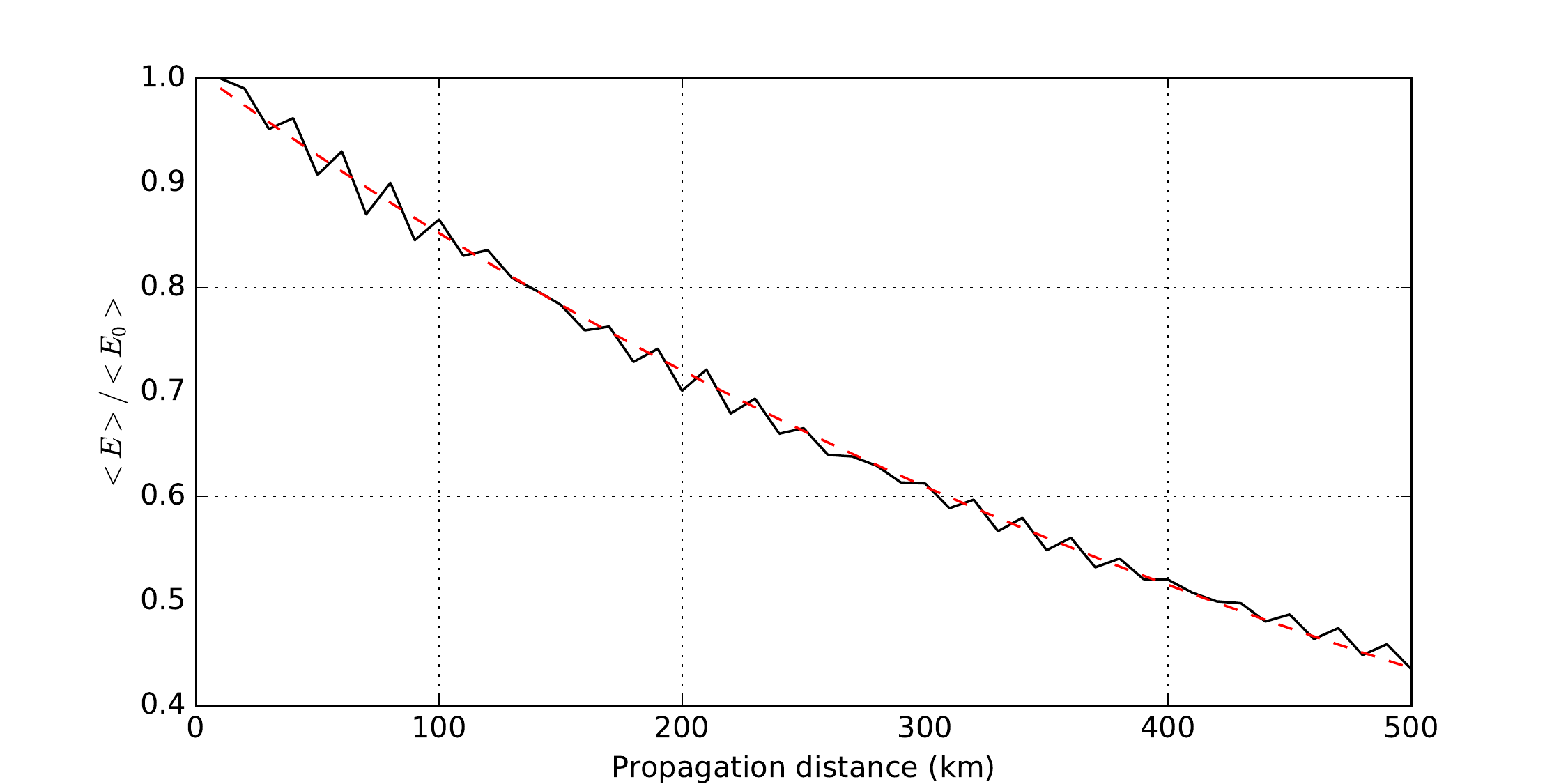}
\caption{Decay of the signal inside the cavity (averaged energy of the large 200ns pulses as a function of the propagation distance inside the $10$-km long cavity. The experimental data are fitted by using $\langle E(z) \rangle/\langle E(0) \rangle=\exp(-\alpha_{eff} z)$ (red dashed line).}
\label{figSup:3}
\end{figure}

By increasing the power of the pump beam at 1450 nm from zero to a few hundred mW, the propagation distance reached by the 200 ns square pulse increases from 20 km to 500 km. In Fig. \ref{figSup:3}, we show the experimental decay of the signal inside the cavity measured from the averaged energ of the 200 ns pulses over 19 realizations. By fitting experimental data using a least square method, we find that effective losses coefficient $\alpha_{eff}$ is around $1.7 \times 10^{-3}$km$^{-1}$.

\subsection{Spatio-temporal diagram}

Extracting the dynamics in space and time from the raw data recorded with the oscilloscope requires a very precise temporal post-synchronization of each 500 ns window (see Fig. 2) corresponding to consecutive round trips.

\begin{figure}[h]
\includegraphics[width=8cm]{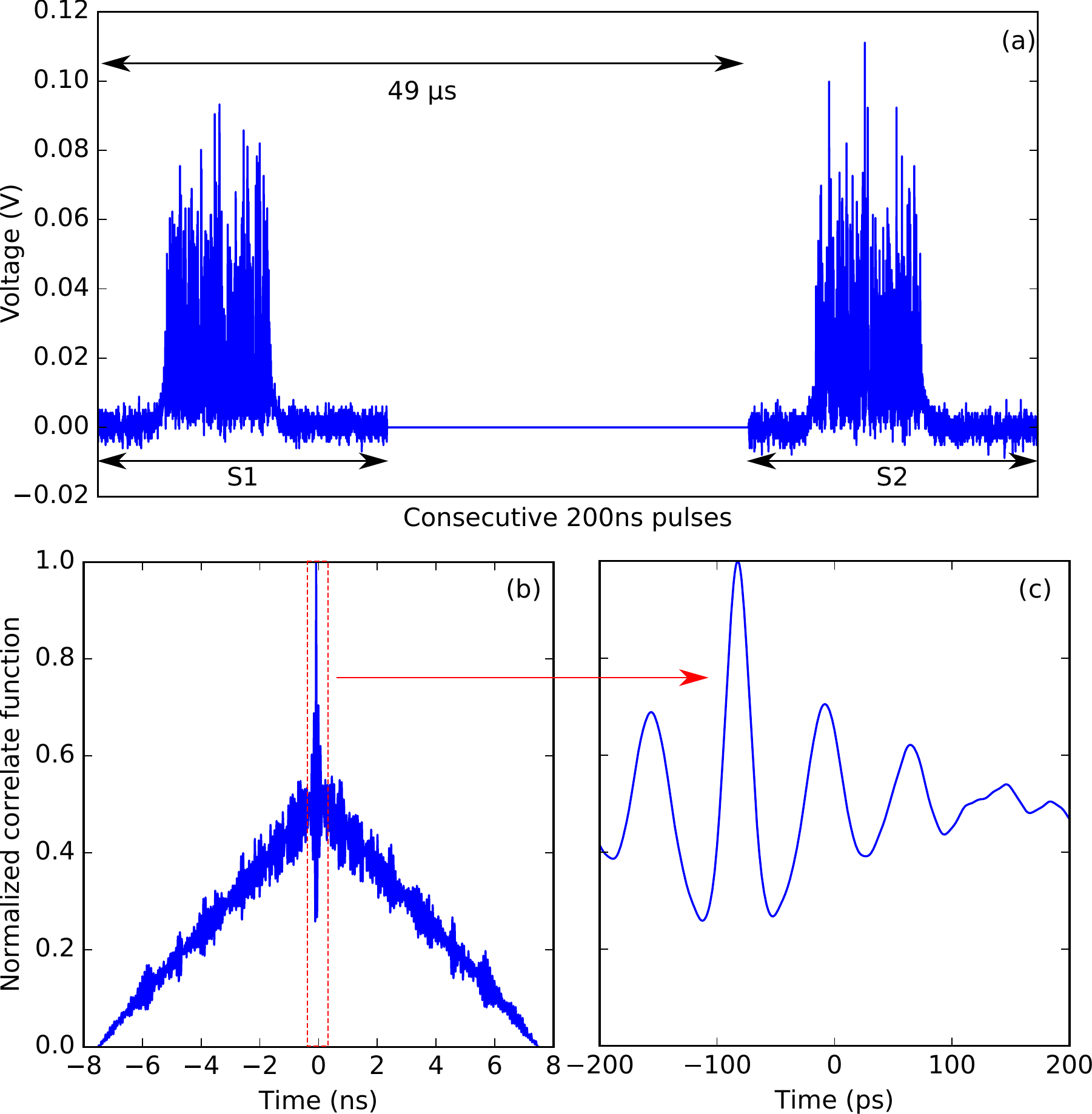}
\caption{(a) Two consecutive time series S1 and S2 with durations of 500 ns. (b) Normalized cross-correlation function between S1 and S2. (c) Zoom on the maximum of the normalized cross-correlation function. }
\label{figSup:4}
\end{figure}

We first compute the cross-correlation function between two consecutive sequences separated by $\sim 49 \mu$s (see Fig. \ref{figSup:4}). The position of the maximum of the cross-correlation function ($-82$ ps in the example of Fig. \ref{figSup:4}) provides the shift that is needed to be applied. Note that the signal is first interpolated in order to achieve  a better resolution of the correlation function.  This scheme is repeated for every consecutive pulses in order to build the full spatio-temporal diagram.

\section{Section II: Influence of losses (numerical simulations of 1D-NLSE)}

In the Letter, we have considered 1D-NLSE with a small linear effective damping. Here, we investigate the role of the losses on the statistics and we compare numerical simulations of 1D-NLSE (\ref{eq:NLSE})that are made with and without losses ($\alpha_{eff} = 0 $).
 
\begin{equation}\label{eq:NLSE_SUP}
i\frac{\partial \psi}{\partial z}=\frac{\beta_2}{2}\frac{\partial^2
  \psi}{\partial t^2}-\gamma|\psi|^2\psi -i \frac{\alpha_{\rm eff}}{2}
\psi.
\end{equation}

Our numerical simulations are performed by using a step-adaptive
pseudo-spectral method. Numerical simulations are performed by using a box
of size L =4000 ps,  with a numerical mesh of 2048 points. Statistical
properties of the continuous wave seeded by noise are computed over an
ensemble of 200 realizations.

\subsection{Dynamics}

\begin{figure}[h]
\includegraphics[width=8.2cm]{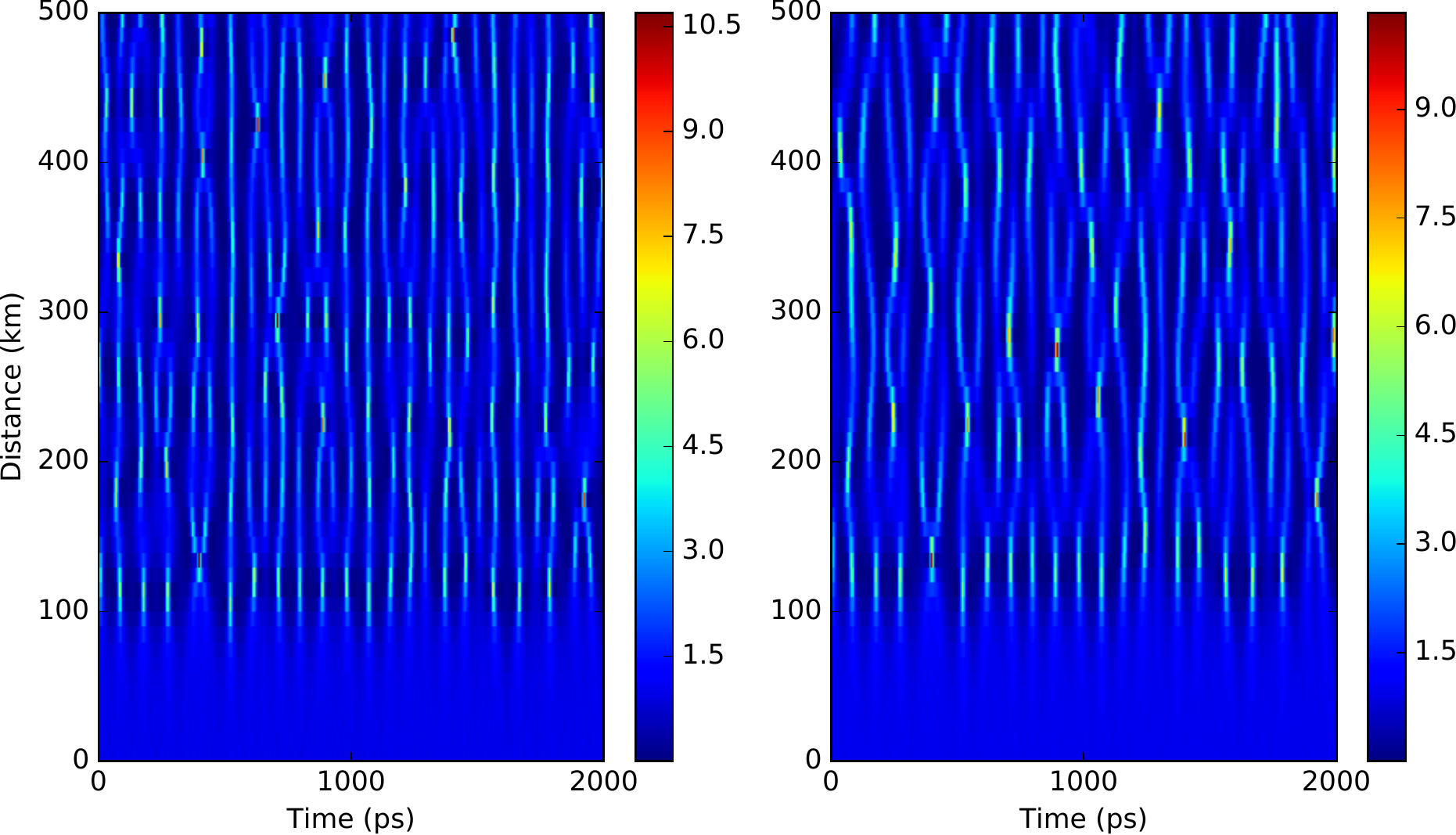}
\caption{(a) Space time diagram of MI without damping, (b) Space time diagram of MI with damping  $\alpha_{eff} = 1.675 \times 10^{-3}$.}
\label{figSup:5}
\end{figure}

Fig. 2(b) in the Letter represents the space-time evolution that is computed by taking into account the effective losses measured in experiments. In  Fig. \ref{figSup:5}, we compare numerical simulations made with and without effective losses. There is no significant difference between the two evolutions even though one can notice a very small enlargement of the breather-like structures during the propagation when losses are taken into account.

\subsection{Second order moment}

Fig. 3 of the Letter shows the evolution of the second order moment of the power, i.e. the fourth order moment of the amplitude of the field  $\kappa_4(z)=   \langle  P(z,t)^2  \rangle / \langle P(z,t)  \rangle^2$  as function of the propagation distance $z$.

\begin{figure}[h]
\includegraphics[width=9cm]{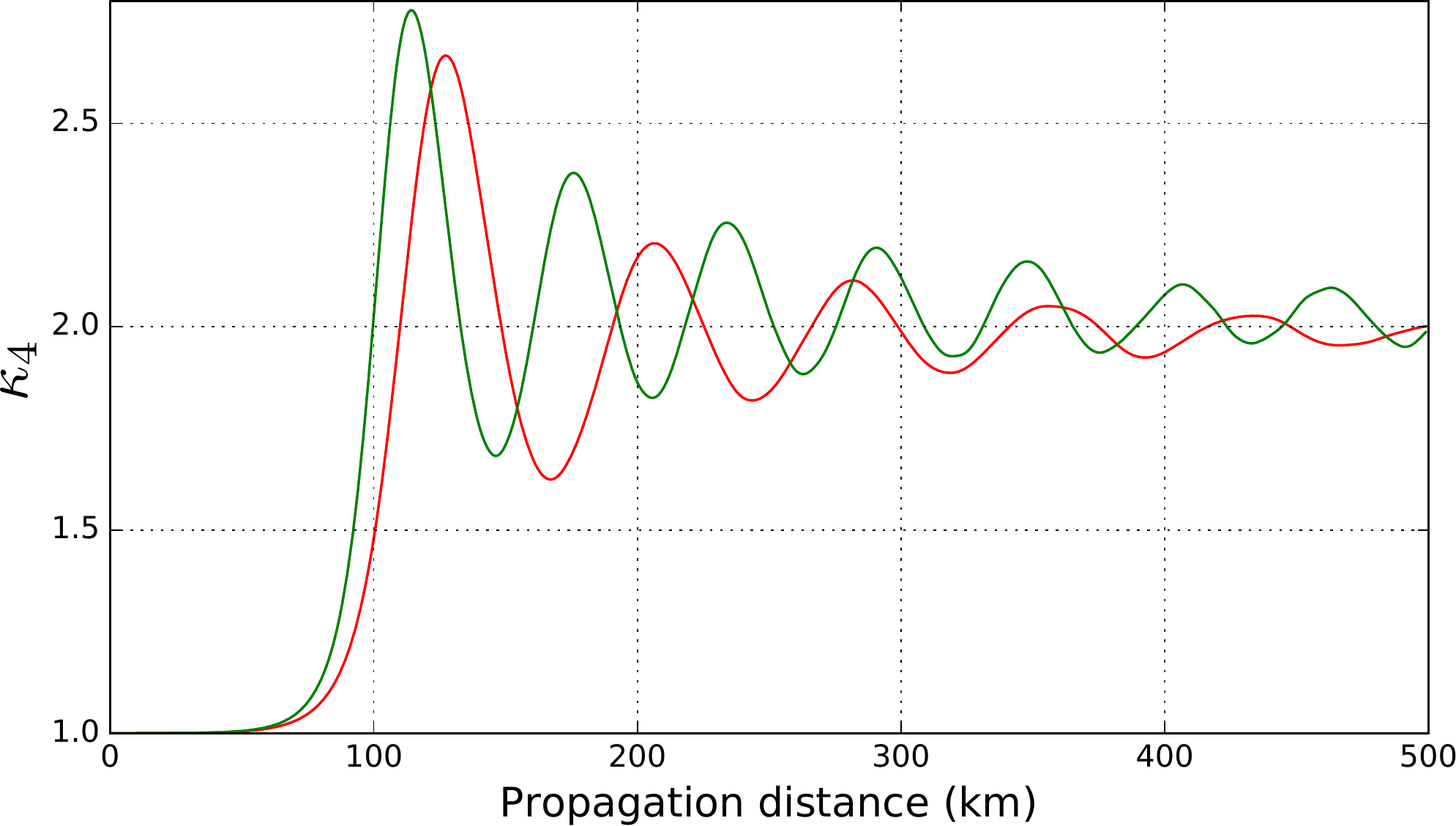}
\caption{Evolution of the moment $\kappa_4$ as function of the propagation distance. In red line, the moment is computed with linear damping $\alpha_{eff} = 1.675 \times 10^{-3}$ km$^{-1}$ whereas the green curve corresponds to numerical simulations made with no losses $\alpha_{eff} = 0$. The same experimental parameters are used in the two numerical simulations : $\beta_2 =-22$ ps$^2$km$^{-1}$,$\gamma = 1.3$W$^{-1}$km$^{-1}$, P$_0 = 43$ mW.  Note that here, the finite bandwidth of detection (see below) is {\it not} included. }
\label{figSup:6}
\end{figure}

 In Fig. \ref{figSup:6},  we compare the evolution of the second order moment of the power computed from numerical simulations with and without linear damping ($\alpha_{eff} \neq 0$ or $\alpha_{eff}=0$)

\subsection{Autocorrelation of optical power (second order of coherence)}

In integrable turbulence ($\alpha_{eff}=0$ in Eq.(\ref{eq:NLSE_SUP}) ) with initial conditions made of a plane wave with noise, the long term evolution is characterized by an exponential distribution of the power $P=|\psi|^2$ corresponding also to a Gaussian single-point statistics of $\psi$. Up to now, no distinction between this nonlinear state and the simple linear superposition of waves (central limit theorem) has been made. In the main Letter, we have shown that the second order autocorrelation function  $g^{(2)}$ provides a clear signature of the nonlinear stage of MI. The autocorrelation function exhibits an oscillatory structure. However, the visibility of the oscillations is significantly reduced by the influence of losses at long distance of propagation. We provide here the comparison between autocorrelation functions of the power computed from numerical simulations in both cases $\alpha_{eff} \neq 0$ and $\alpha_{eff}=0$. Note that, here, we do not apply the filtering effect induced by the finite detection bandwidth (see Sec. III below).

\begin{figure}[h]
\includegraphics[width=8cm]{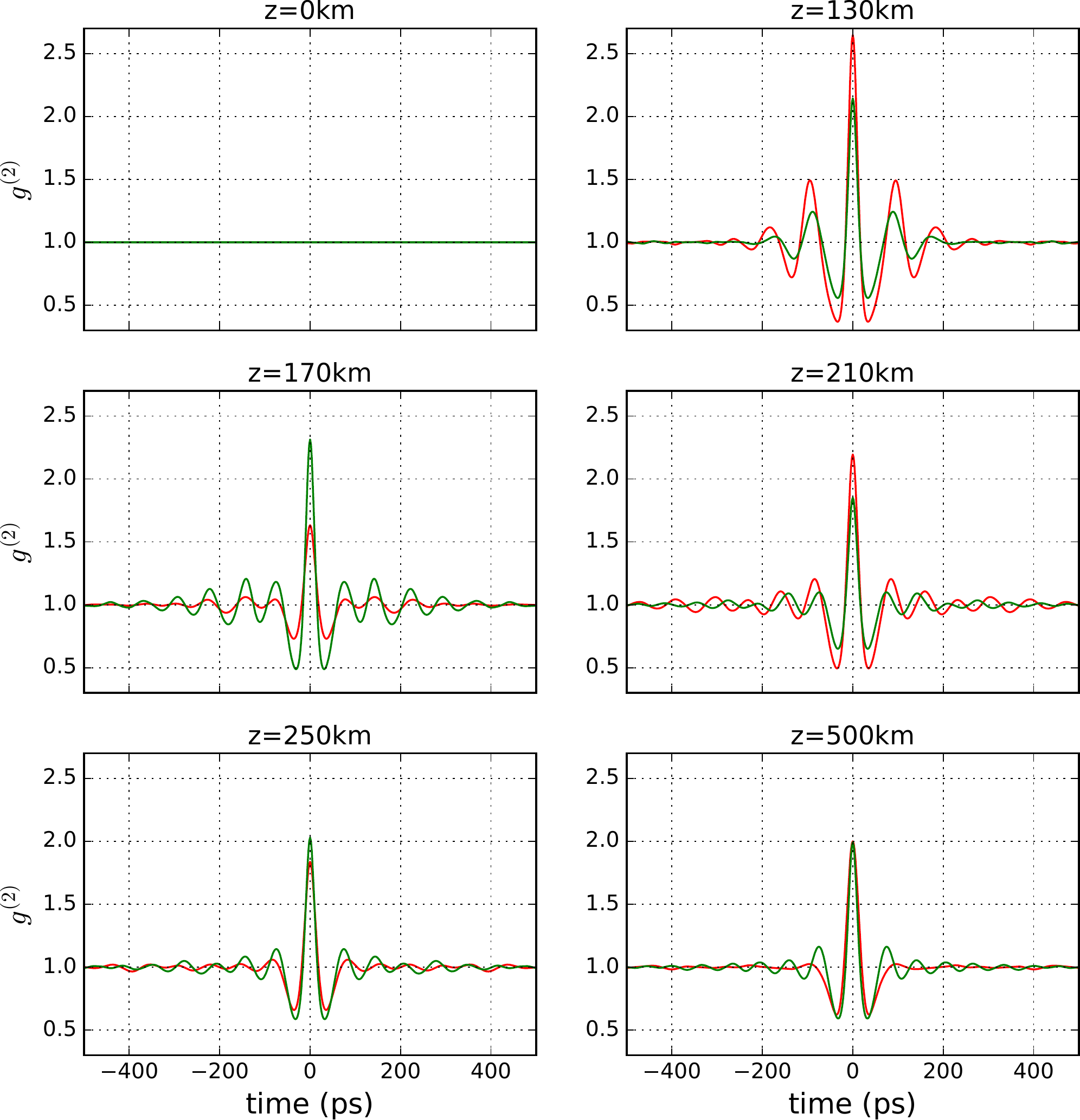}
\caption{Autocorrelation of the power with losses ($\alpha_{eff} = 1.675 \times 10^{-3}$ km$^{-1}$, red line) and without losses  ($\alpha_{eff} = 0$ km$^{-1}$, green line). Other parameters used in numerical simulations are : $\beta_2 =-22$ ps$^2$km$^{-1}$,$\gamma = 1.3$W$^{-1}$km$^{-1}$, P$_0 = 43$ mW.}
\label{figSup:7}
\end{figure}

The comparison reveals that the effective losses play a significant role on the quasi-periodic structure of $g^{(2)}$ at long propagation distance. The oscillations of $g^{(2)}$ gradually disappear in the presence of effective damping while they remain observable in the integrable case for $z\ge 200$ km.

\section{Section III: filtering effect induced by the finite detection bandwidth}

\subsection{Influence of the bandwidth of detection}

In experiments, the dynamics is monitored by fast photodiodes (Picometrix D-8IR), having a nominal bandwidth at $-3dB$ of 65 GHz, connected to a fast oscilloscope (LeCroy Labmaster 10-65ZI) having a nominal bandwidth  $-3dB$ of 65GHz and a sampling rate of 160 Gsamples/s. In the letter, we report simulations in which the bandwidth of detection is modeled by a simple first order filter. We provide here the procedure used to evaluate the bandwidth of the effective first order filter.

\begin{figure}[h]
\includegraphics[width=7cm]{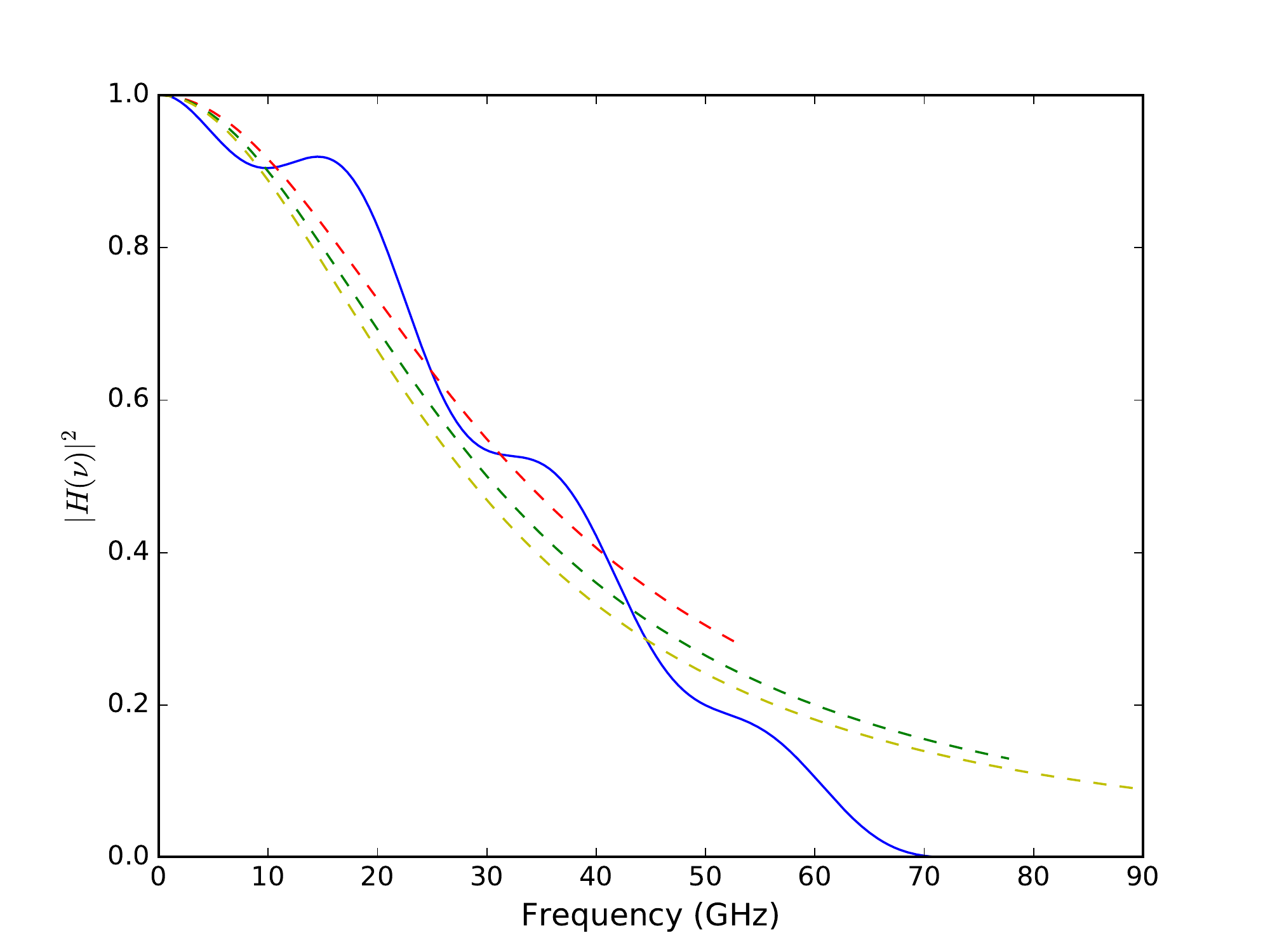}
\caption{(a) Modulus square of the experimental transfer function $|H(\nu)|^2$ (blue line), low-pass fit made with different frequency windows : 53 GHz (dashed yellow lines), 78 GHz (dashed green line), full frequency window (red dashed red line).}
\label{figSup:8}
\end{figure}

 By using  picosecond pulses emitted by a mode-locked fiber laser, we have measured the impulse response of our detector (photodiode + oscilloscope). We report in Fig. \ref{figSup:8} the modulus of the Fourier transform of the impulse response, i.e. the modulus square of the transfer function $|H(\nu)|^2$ of our detection apparatus. We report in  \ref{figSup:8} different adjusted fits that can be made with a first order :

\begin{equation}
|H(\nu)|^2 = \Big |\frac{1}{1+i \frac{\nu}{\nu_c} } \Big |^2
\label{Eq:filtre}
\end{equation}

The cut-off frequency given by the best adjustement (least-squares) procedure depends on the frequency window on which the adjustement is performed. The value of the cut-off frequency  is typically found between $\nu_c \sim 28$ GHz and $\nu_c \sim 33$GHz  (Fig. \ref{figSup:8}). We discuss now  the influence of this finite bandwidth effect by applying the low pass filter described by Eq. (\ref{Eq:filtre}) with $\nu_c = 28$ GHz to the optical power computed with numerical simulations of 1DNLSE.

\begin{figure}[h] 
\includegraphics[width=8cm]{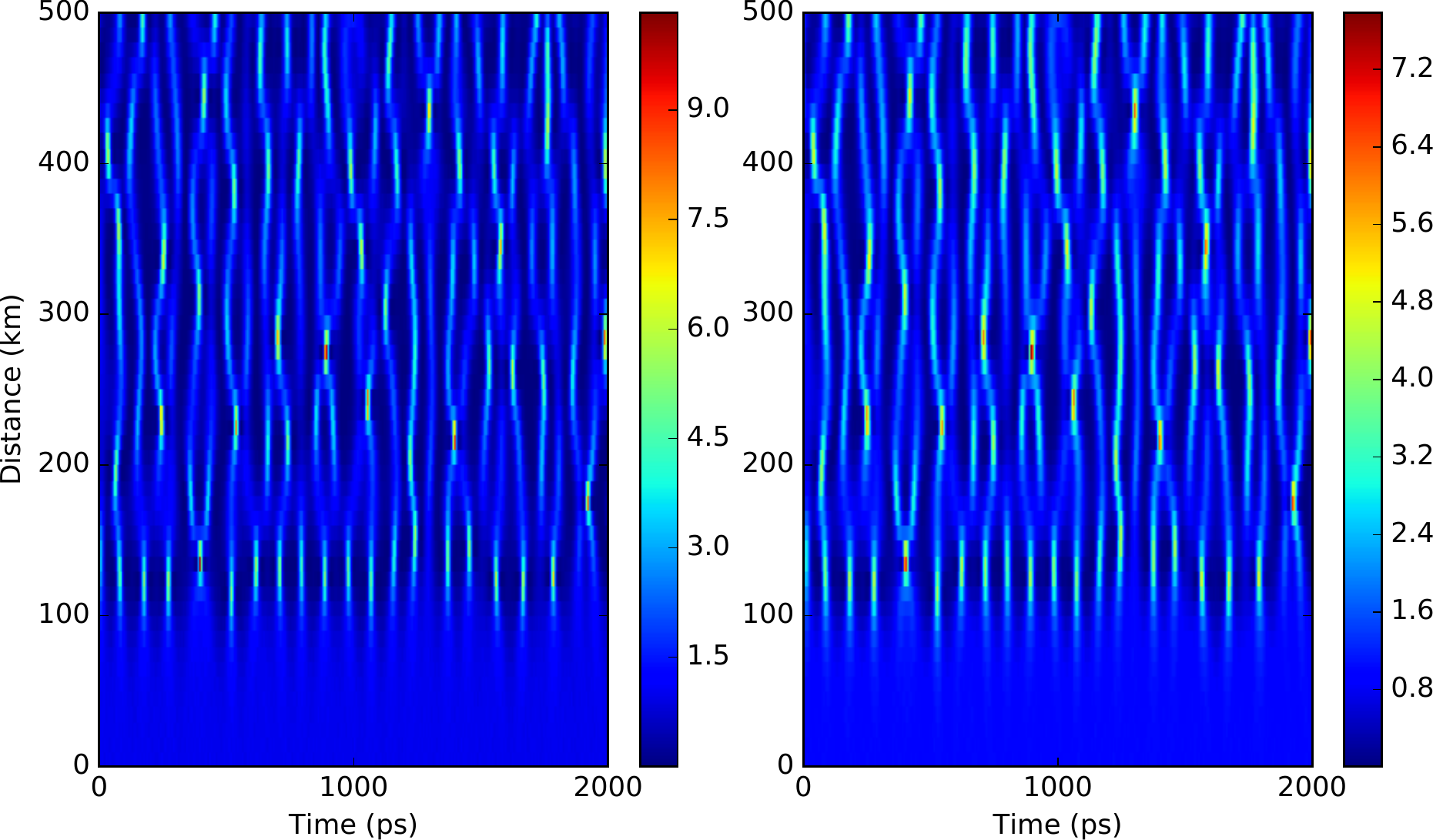}
\caption{{\bf Numerical simulation of the modulation instability}. {\bf(a)} Space time diagram of MI without filtering effect.{\bf(b)} Space time diagram of MI by taking into accound the low-pass filter effect with $\nu_c=28$~GHz.}
\label{figSup:9}
\end{figure}

\begin{figure}[h]
\includegraphics[width=7cm]{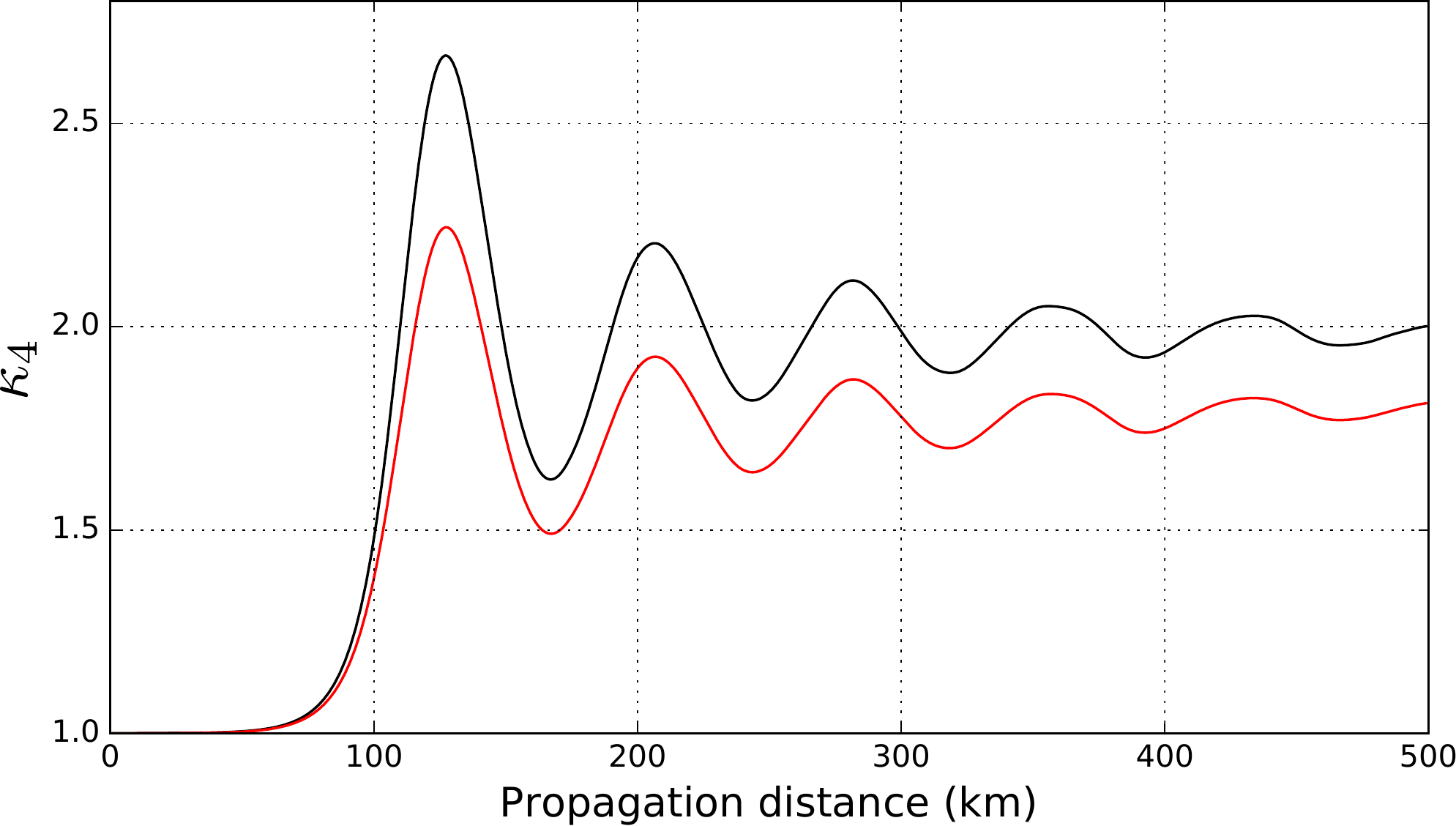}
\caption{Evolution of the moment  $\kappa_4(z)$ with the propagation distance (numerical simulations). The integrable case is plotted in black line while the numerical results filtered with $\nu_c=28GHz$ is plotted in red line. The parameters are $\beta_2 =-22$ ps$^2$km$^{-1}$,$\gamma = 1.3$W$^{-1}$km$^{-1}$, P$_0 = 43$ mW, $\alpha = 1.675 \times 10^{-3}$ km$^{-1}$.}
\label{figSup:10}
\end{figure}

The Fig. \ref{figSup:9} demonstrates that the observed dynamics is only slightly influenced by the limited detection bandwidth of our experiments. The finite detection bandwidth has some impact on the stationary value reached by the second order moment of the power $\kappa_4$. The stationary value computed in the integrable case is  $k_4= 2$ while the value observed in the exepriments is $k_4\sim 1.8$. Fig. \ref{figSup:10} displays the comparison of the evolution of $\kappa_4(z)$ computed in numerical simulations of 1DNLSE by taking into account the effect of bandwidth of detection or not. One immediately sees that the detection bandwidth influences the stationary state value reached by $\kappa_4$ but that it almost does not influence the quasi-periodic transient regime of the evolution of $\kappa_4$.

\begin{figure}[h]
\includegraphics[width=7cm]{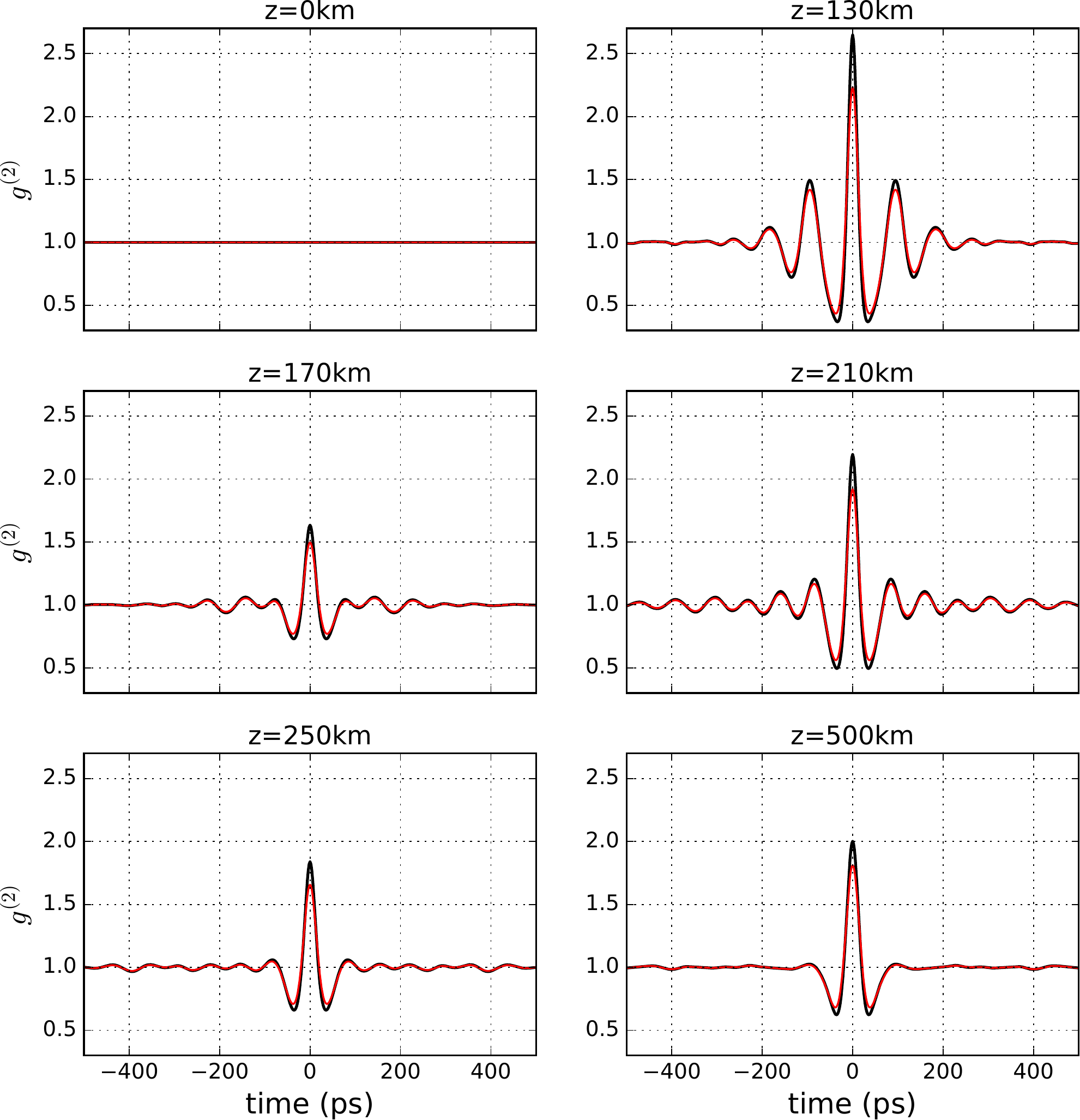}
\caption{Autocorrelation function $g^{(2)}$ for several propagation distances, without filtering effect (in black line) and with low pass filter associated to the  bandwith of detecion equal to $28$ GHz (in red line).  Numerical simulation of 1DNLSE are made with $\beta_2 =-22$ ps$^2$km$^{-1}$,$\gamma = 1.3$W$^{-1}$km$^{-1}$, P$_0 = 43$ mW, $\alpha = 1.675 \times 10^{-3}$ km$^{-1}$.}
\label{figSup:11}
\end{figure}

We finally compare the autocorrelation $g^{(2)}(\tau)$ computed from numerical simulations by taking into account or not the bandwidth of detection (see Fig. \ref{figSup:11}). One immediately sees that the value $g^{(2)}(0)$ is reduced by the bandwidth of detection in our experiments while the oscillatory structure is almost not affected by the detection process. As a conclusion, the precise oscillatory characteristics of $g^{(2)}$ are mainly influenced by losses and not the finite detection bandwidth in our experiments. On the contrary, the value at $\tau=0$ of $g^{(2)}$ is mainly influenced by the effective losses in the fiber loop.

\section*{Acknowledgments}

This work has been partially supported by the Agence Nationale de la
Recherche through the LABEX CEMPI project (ANR-11-LABX-0007)   and by
the Ministry of Higher Education and Research, Nord-Pas de Calais
Regional Council and European Regional Development Fund (ERDF) through
the Contrat de Projets Etat-R\'egion (CPER Photonics for Society
P4S). The work of D.A. (simulations) was supported by the state assignment of IO RAS, theme 0149-2019-0002.


%

\end{document}